\documentclass[letterpaper, 10 pt, conference]{ieeeconf} 

\IEEEoverridecommandlockouts
\usepackage{tabularx, graphicx,
  amsmath,amsfonts,siunitx,pifont,amssymb}
\usepackage[table]{xcolor}
\usepackage{xcolor, bm, float, multirow, multicol, courier, mathtools, url}
\usepackage[ruled,vlined]{algorithm2e} \pdfminorversion=4
\let\labelindent\relax \usepackage{enumitem} 
\usepackage{adjustbox}
\usepackage{subcaption}
\usepackage{todonotes}
\usepackage{booktabs}
\usepackage{makecell}
\usepackage[protrusion=false]{microtype}
\usepackage{tensor}
\usepackage[bottom]{footmisc}
\usepackage{tikz}
\usetikzlibrary{bayesnet}
\usetikzlibrary{arrows,positioning}

\newcommand{\etal}{\textit{et al}.~}

\usepackage{pdfpages}
\usepackage{cite}
\usepackage{balance}
\usepackage{hyperref}

\newcommand{\milad}[1] { \textcolor{blue}{MR: #1}}

\newcommand\norm[1]{\left\lVert#1\right\rVert}

\def\secref#1{Sec.~\ref{#1}}
\def\figref#1{Fig.~\ref{#1}}
\def\tabref#1{Table~\ref{#1}}
\def\eqref#1{Eq.~(\ref{#1})}
\def\algref#1{Alg.~\ref{#1}}

\title{\LARGE \bf{Online Estimation of Diameter at Breast Height (DBH)\\ of Forest Trees Using
a Handheld LiDAR}}

\author{Alexander Proudman\textsuperscript{1}, Milad Ramezani\textsuperscript{1} and Maurice
Fallon\textsuperscript{1}
    \thanks{This research is supported by the UKRI/ESPRC ORCA Robotics Hub (EP/R026173/1). M. Fallon is supported by a Royal Society University Research Fellowship.}
    \thanks{\textsuperscript{1} These authors are with the Oxford Robotics Institute, University of Oxford, UK.
        {\tt\small alexander.proudman@wadham.ox.ac.uk, \{milad, mfallon\}@robots.ox.ac.uk}}%
}


\begin{document}
	
\setlength{\abovedisplayskip}{4pt}
\setlength{\belowdisplayskip}{4pt}
	
\maketitle 
\thispagestyle{empty} 
\pagestyle{empty}
	

\begin{abstract}
While mobile LiDAR sensors are increasingly used to scan in ecology and forestry applications,
reconstruction and characterisation are typically carried out offline (to the best of our
knowledge). Motivated by this, we present an online LiDAR system which can run on a handheld device to segment and track
individual trees and identify them in a fixed coordinate system. Segments relating
to each tree are accumulated over time, and tree models are completed as more scans are captured from
different perspectives. Using this reconstruction we then fit a cylinder model to
each tree trunk by solving a least-squares optimisation over the points to estimate the Diameter at Breast Height (DBH) of the trees. 
Experimental results demonstrate that our system can estimate DBH to within $\sim$7 cm accuracy for 90\% of individual trees in a forest (Wytham Woods, Oxford).
\end{abstract}
	
\section{Introduction}

Researchers in ecology and forestry monitor the size and growth of individual trees within a forest to infer arboreal health. It is common to retrieve quantitative metrics such as Diameter at Breast Height (DBH) using manual methods, such as measuring tapes, to ensure consistency with previous surveys.

More recently approaches using 3D LiDAR scanners have been presented for tree segmentation and the construction
of Quantitative Structure Models (QSMs)~\cite{raumonen2013fast}. These models can be used to
estimate the above-ground biomass and carbon stock by estimating the volume of individual
trees~\cite{calders2015nondestructive,gonzalez2018estimation}. However, retrieval of
these tree-scale metrics requires data processing of a large-scale point cloud which is often
conducted offline in existing approaches.

In this paper, we present an online point-cloud processing tool chain which can
segment and track individual trees relative to a fixed coordinate frame. The ground surface can also
be inferred, allowing us to estimate the DBH of each tree (at 1.4m height) as the scanning process goes on.

\begin{figure}[t!]
\includegraphics[width=0.5\textwidth]{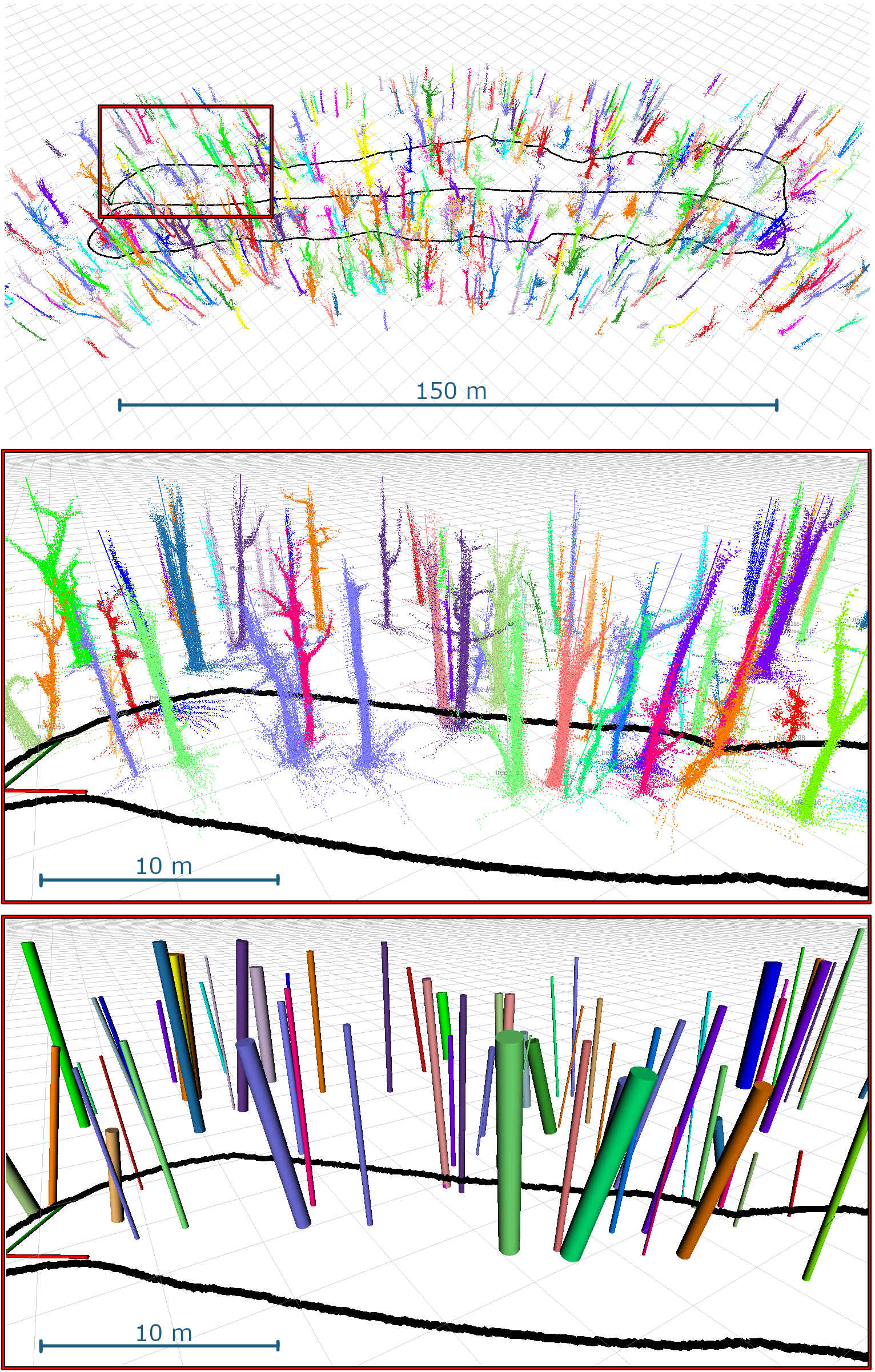}
\caption{\small{Visualisation of the system output.
\textbf{Top}: a high level view showing a 150 m transect. The path taken by the person carrying the
sensor in
\figref{fig:map} is shown in black. \textbf{Centre} and \textbf{Bottom}: Detail of a section is
shown with two
different representations: one showing each tree segmented and individually coloured and the other shows a cylinder (primary axis and diameter) fitted at breast height.}}
\label{fig:cover_fig}
\vspace{-5mm}
\end{figure}

The contributions of this work are as follows:
\begin{itemize}
     \item An online system to track trees in dense forests and to identify individual trees.
     \item An experimental evaluation, of our system, in a forest area and the evaluation of the tree DBH using our system. Results are compared to measurements taken manually at the test site as well as comparison against a commercial tool (using the final accumulated point cloud).

\end{itemize}

\textbf{Video:} \url{https://youtu.be/Ut7d2eOypko}

The remainder of this paper is organised as follows. In Section
\ref{sec:RelatedWork} we discuss
related work. Section \ref{sec:Methods} explains how we designed our methods. 
Experimental results are presented in Section 
\ref{sec:Experiments}. Section \ref{sec:FutureWork} discusses conclusions
and suggests some topics for future work.

\section{Related Work}
\label{sec:RelatedWork}
Advances in LiDAR technology have led to it being used in non-traditional fields such as ecology and
forestry. Both stationary terrestrial laser scanning and portable scanning LiDAR have been
used. In this section
we review the approaches that utilise LiDAR for individual tree segmentation.

\subsection{Tree Segmentation Using Terrestrial LiDAR}
Many existing tree segmentation techniques operate on a single unified point cloud collected by a
Terrestrial Laser Scanner (TLS) from a small number of static scanning stations in post-processing. Raumonen~\etal~\cite{raumonen2015massive} and
later Trochta~\etal~\cite{trochta20173d} clustered these scans into smaller point clouds to
work around large size of these scans and to reduce memory consumption. They then segmented
tree-level clouds within each cluster by growing segments using assumptions of fixed inter-cluster distance and orientation to infer connectivity. Methods such
as~\cite{zhong2016segmentation} benefited from graph theory to find the connectivity between
adjacent points and to finally segment each tree. These approaches, however, rely on multiple
assumptions about a tree's architecture as well as assuming minimal
interconnection between their crowns.

Recently, Burt~\etal~\cite{burt2019extracting} presented a software package named \textit{treeseg}
in which a region-growing technique is used to segment individual trees. Key features in
\textit{treeseg}'s design are its independence of forest type and instrument, and assumptions about the
tree structures. However, \textit{treeseg} and the techniques aforementioned were
mostly tested with high quality data
collected by stationary
laser scanners. The point clouds of these scanners were very accurate with minimal noise and therefore tree
individuals were easy to identify.

There are also some techniques
to automatically segment trees utilising learning
pipelines such as a trained random forest classifier~\cite{digumarti2018automatic} or Convolutional
Neural Networks (CNNs)~\cite{digumarti2019approach}, however these were applied offline using data
collected from individual trees.

\subsection{Tree Segmentation Using Mobile LiDAR}

Although mobile LiDAR systems, which are normally carried by an agent at walking pace, make
ranging measurements with lower precision than TLS, they have drawn attention in
environmental science because of the falling cost of the sensor and the quick data collection time.
Heo~\etal~\cite{heo2019estimating} used mobile LiDAR to collect data in an urban area, including parks and streets, to estimate the height of trees and their DBH by calculating the height-above-ground and
using a least-squares circle fit approach~\cite{pratt1987direct}. They emphasised the advantage of
using mobile LiDAR to reduce shadow/occlusion effects, which are more prominent with terrestrial
LiDAR systems, especially in an urban environment. They utilised a Stencil LiDAR system
produced by
Kaarta\footnote{https://www.kaarta.com/} which includes a Velodyne sensor.

Similarly, Zhou~\etal~\cite{zhou2019extracting} collected LiDAR data with a Velodyne VLP-16 LiDAR
sensor. In offline processing the authors estimated the DBH of the trees. They removed
the points residing on the ground and estimated the DBH using
Random Sample Consensus (RANSAC) algorithm on the segments produced by an Euclidean-based
clustering algorithm~\cite{trevor2013efficient}.

Westling~\etal~\cite{westling2020graph} scanned individual avocado trees with a 5 m spacing using a
GeoSLAM Zebedee 1 handheld device. The authors first voxelised point clouds and conducted a
graph-based search over the voxels to find all paths connecting to a root voxel, which was considered
to be the tree node. The tree node is segmented from the ground by comparing the height of points
locally within a search radius.

These approaches have been tested after data collection which doesn't allow the user to perceive the reconstruction during scanning.
Our
motivation is to develop a LiDAR-driven technique which can estimate the structural parameters of
individual trees in dense forestry areas with rough terrain in real-time. A real-time mapping system can
ensure full coverage by providing feedback to the operator, so that gaps are not left, ensuring the environment is fully scanned and processed at runtime.

\begin{figure}[t]
    \centering
    \includegraphics[width=0.8\linewidth]{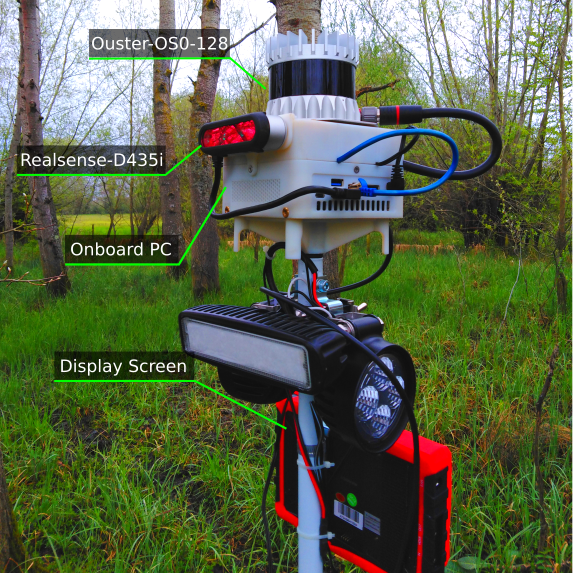}
    \caption{\small{The handheld device used in our system. The labelled components work together
    to provide online feedback on the forestry data collection}}
    \label{fig:system}
\end{figure}

\section{Methods}
\label{sec:Methods}
%

As shown in~\figref{fig:system}, our device consists of an Ouster OS0-128 LiDAR and an Intel
Realsense D435i, each with a
built-in IMU. However, we only use the Realsense D435i for video in this system.
The LiDAR has 128 beams and a $90^{\circ}$ vertical field of view.
While it cannot completely cover the environment vertically, the $90^{\circ}$ vertical field
is sufficient to identify the individual trees within about 20 m of the sensor.
The sensors used in systems such as Zeb-Revo or Zebedee\footnote{https://geoslam.com/} have narrow field of view
and require the sensor to be either rotated or oscillated to cover the environment entirely.

\begin{figure}[t]
    \centering
    \includegraphics[width=1.0\linewidth]{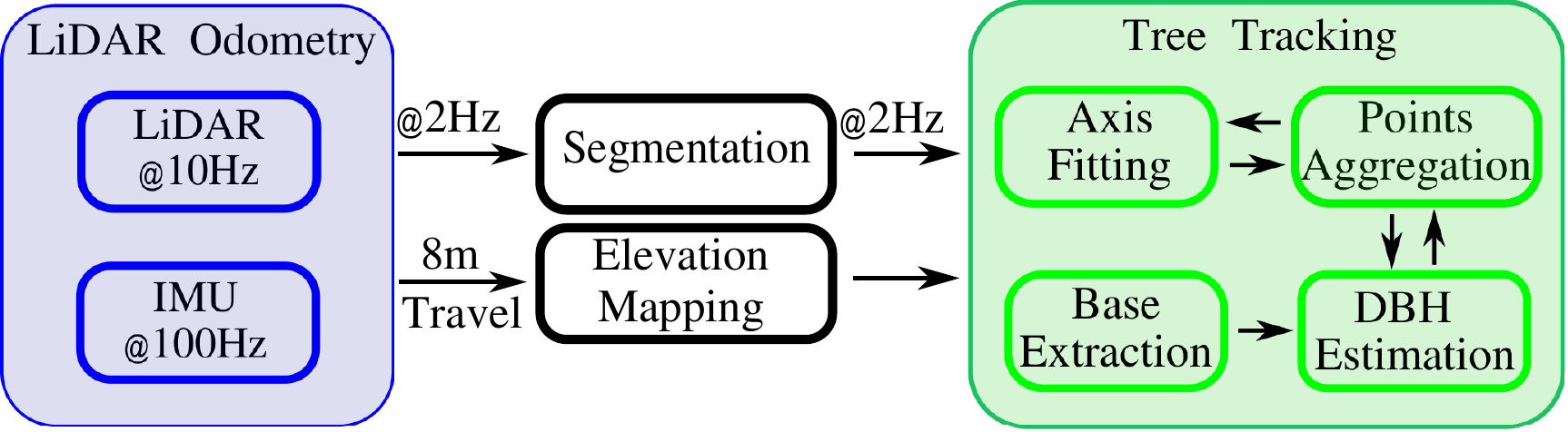}
    \caption{\small{An overview of the system pipeline, the frequencies of communication between
blocks are given. Elevation mapping is triggered every 8 m travelled.}}
    \label{fig:system_overview}
\end{figure}

Our proposed system architecture is shown in \figref{fig:system_overview}. Due to the system's
real-time nature, the pipeline is different from many existing methods, and alterations have been
made to parallelise and speed up the overall pipeline.
For example our system builds up an elevation map rather than removing ground points prior to segmentation as done by Heo~\etal~\cite{heo2019estimating} and Sanzhang~\etal~\cite{zhou2019extracting}. Removing the ground points improves the segmentation of trees. This has the potential to impede the real-time operation of our system, see \secref{sec:Experiments} for demonstration of this. Our system develops counter-measures to the potentially poorer segmentation detailed throughout this section specifically \secref{sec:LoBF}.


In the following, we describe the algorithms and methods used by each functional block within the
system architecture. Most methods used have been proven to work for terrestrial LiDAR in
post-processing, only requiring significantly tall trees such that the trunk is clearly separate from the canopy. We use a LiDAR-inertial odometry module to accumulate a point cloud reconstruction of the environment around the sensor. The \textit{Tree Tracking} block, not used in the typical LiDAR
forest inventory systems, segments and processes each tree detection. Advantages of using this online fused data approach are that:

\begin{itemize}
     \item Previous scans, recorded using this system, can be used to instantly identify trees while in the field.
     \item Success of the data collection can be evaluated online, including the sensor's coverage.
     \item Data collection and analysis only takes the amount of time needed to walk along a selected transect (a narrow section of land along which measurements are taken).
\end{itemize}


\subsection{LiDAR Odometry}
\label{sec:LiDAR_Odometry}

Our LiDAR odometry system~\cite{wisth2019robust} is a factor-graph based windowed smoother which
fuses Inertial Measurement Unit (IMU) readings with measurements from a multi-beam LiDAR scanner
mounted on a handheld
device (\figref{fig:system}). The particular configuration used in this work, uses IMU
pre-integration to remove motion distortion from scans and to initialise Iterative Closest Point (ICP) registration, specifically the
implementation of Pomerleau~\cite{pomerleau2013applied}, to determine the relative
transform for every 2 metres of distance travelled.

The IMU measurements and the relative
transforms then form constraints in a factor graph to estimate the poses in a sliding window
optimisation utilising iSAM2~\cite{kaess2012isam2} (as part of the GTSAM library). Additionally,
since IMU provides gravity information, we can align the clouds with the gravity direction. This alignment
helps determine the base of trees using the $z$ direction. Evaluation of our LiDAR odometry system is
discussed in~\secref{sec:lidar_odom}.



\subsection{Segmentation}
\label{sec:Segmentation}

To roughly extract the tree trunks, branches, canopy and shrubbery in the original point cloud,
Euclidean segmentation is carried out. Using the PCL library\footnote{https://pointclouds.org/},
the cloud is first downsampled using a voxel filter to lessen the radial variation of the point
density and to speed up the segmentation process as also done by \cite{burt2019extracting}.

After voxel filtering the point cloud, it is reformulated into a \textit{k}-d tree to quickly find the nearest neighbours of every point. These points are clustered together into groups of points which are
within
some distance threshold of each other.

Note that the parameters of voxel size, used in filtering, and the threshold distance, used in Euclidean segmentation,
are tuned based on the LiDAR sensor's characteristics and tree sizes. This calibration was done manually, maximising the size of point clusters containing individual trees, to increase the system speed and accuracy.


\subsection{Elevation Mapping}
\label{sec:Elevation_Mapping}
To perform \textit{Base Segmentation} of the trees being tracked, we employ an open-source sensor-centric elevation
mapping framework\footnote{https://github.com/ANYbotics/elevation\_mapping} which is built upon a
universal Grid Map library~\cite{fankhauser2016universal}. The algorithm, presented in~\cite{fankhauser2018probabilistic}, generates consistent elevation maps of
the environment surrounding the sensor, at the same frequency of the LiDAR output (10 Hz). In this work, we use a resolution of 16 cm for the
grid of $32 \text{m} \times32 \text{m}$. 


Nevertheless, due to the presence of foliage and branches, the terrain created by the elevation
mapping software can contain phantom spikes, making it difficult to detect the precise base of certain trees. To solve this problem, we compute the slope of each grid cell. Cells with slope
greater than a threshold are removed. This removes the majority of spikes, resulting
in smooth terrain with some remaining holes. Finally, we utilise a morphological closing filter on the
grid map to fill the holes. 

\begin{figure}[t]
    \centering
    \includegraphics[width=0.8\linewidth, height=0.8\linewidth]{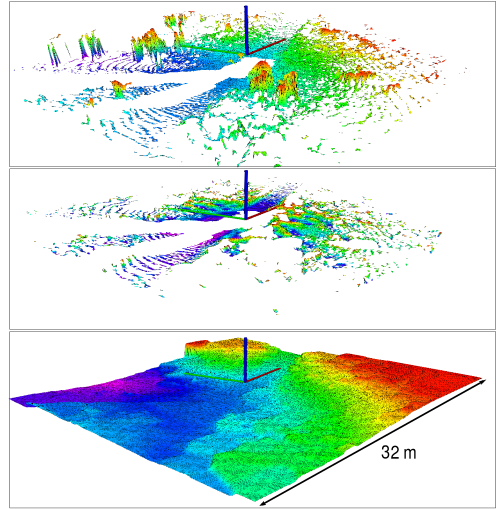}
    \caption{\small{Filtering process on the elevation map: Original elevation map (top) is
filtered to remove spikes (middle). Finally, holes are filled using a morphological closing filter on the grid map
(bottom).}}
    \label{fig:elevation_map}
     \vspace{-5mm}
\end{figure}

Applying the chain of filters is relitievly time consuming so we update the elevation map every 8 m of travel distance based on the state estimate from our LiDAR odometry. This travel distance allows online operation. We analyse the computation time of each
component including elevation mapping in the next section.
\figref{fig:elevation_map} shows this process.


\subsection{Tree Tracking}
\label{sec:Tree_Tracking}

To facilitate a formal definition of a tree the following minimal, `tree descriptor' \(t\) was
defined as follows:

\begin{itemize}
     \item A unique id,
     \item Major axis of the tree incline (\(\textbf{I}\) a vector),
     \item Diameter at Breast Height (DBH) (\(D\)),  
     \item 3D position of the tree base (\(\textbf{b}\)), 
     \item The minimum and maximum height of the point cloud defining the tree (\( min \text{ }\& \text{ } max\)). 
\end{itemize}



Every tree descriptor is derived from a corresponding point cloud of the tree points \(P\) (\( t =
f(P)\)). These points are accumulated over time, gathering information from different angles as
well as reducing the effect of occlusion and improving the estimate of the tree descriptor as~
Heo~\etal~\cite{heo2019estimating} indicated.


Most of the descriptors characteristics are easily evaluated with the exception of the major axis of
the tree.  We used fitting of a line, to the set of tree points (\(P\)), to find this axis.
The algorithm is designed to ignore branches and any potentially segmented crown or ground points
and is described in \secref{sec:LoBF}.

Internally, the \textit{Tree Tracker} holds an inventory of \(n\) distinct trees \(T\) defined by a
descriptor and set of tree points \( T_i \sim t_i,P_i \) for \(0 \leq i \leq n\).


Referring to the system diagram, \figref{fig:system_overview}, incoming clusters \(C\) from the
\textit{Segmentation} block are either discarded or assigned to a new/existing tree descriptor. If
assigned to an existing descriptor, the cluster of points is merged with the corresponding set of tree points and the descriptor is re-evaluated \(P_{new} = P_{old} \bigcup C\), \(t_{new} = f(P_{new})\).

An assign/discard decision is made by evaluating a tree descriptor for each input cluster. A
cluster is confirmed and subsequently assigned only if it has characteristics within the tolerances
of given parameters. The manually adjusted parameter tolerances that resulted in the most consistent tree assignment are:

\begin{itemize}
     \item Requiring the major axis of the tree to be close to vertical, i.e. \(|
\textbf{I}\cdot[0,0,1] | <
\theta_{\text{threshold}} \)
     \item Requiring a minimum height, i.e. \( max-min > h_{\text{threshold}} \)
\end{itemize}

\begin{figure}[t]
    \centering
    \begin{subfigure}{0.46\linewidth}
        \centering
        \includegraphics[width=1.0\linewidth,trim={13cm 0cm 13cm
0cm},clip]{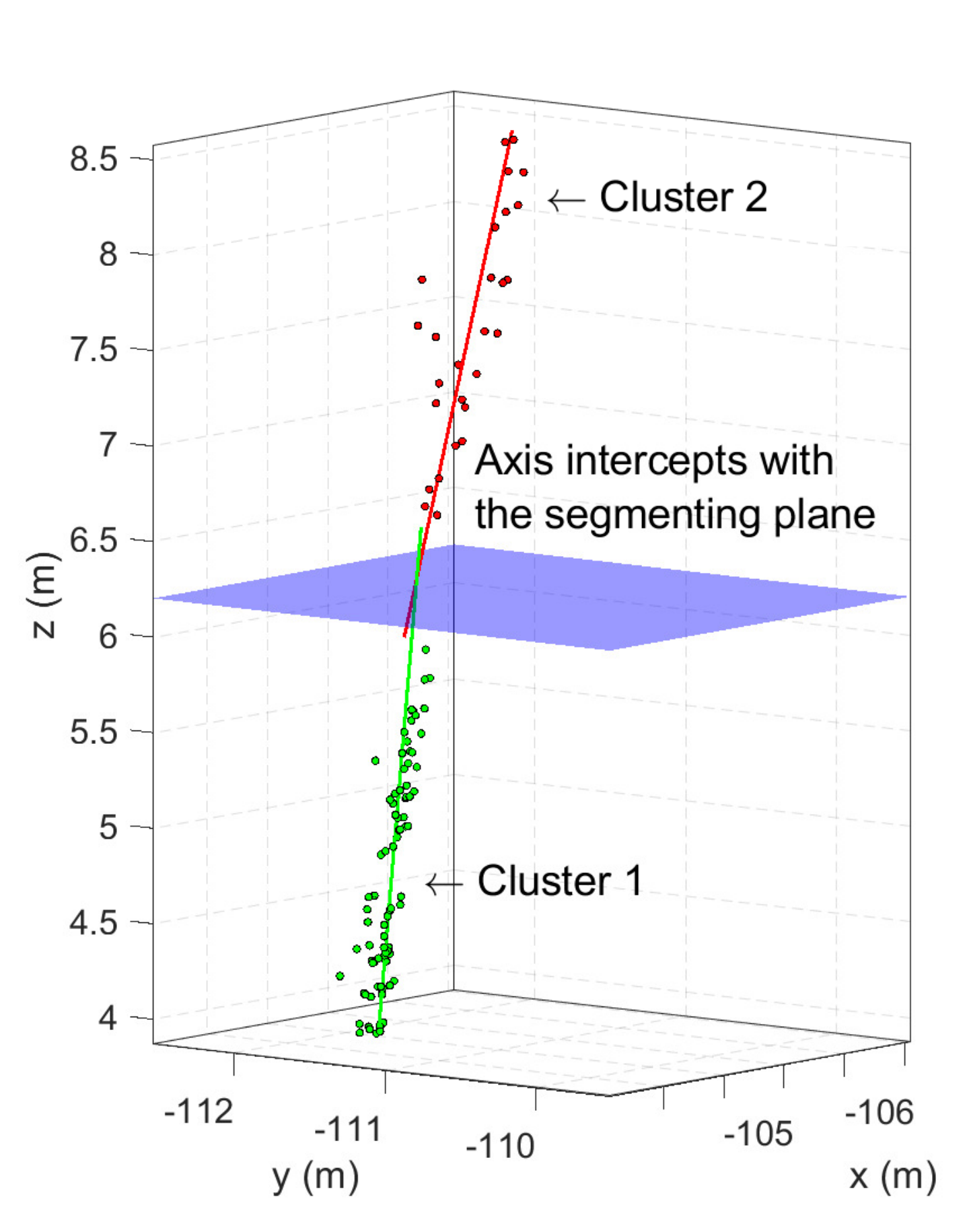}
                \label{fig:tree_axis}
    \end{subfigure}\
    \begin{subfigure}{0.52\linewidth}
        \centering
        \includegraphics[width=1.0\linewidth,trim={12cm 0cm 12cm 0cm},clip]{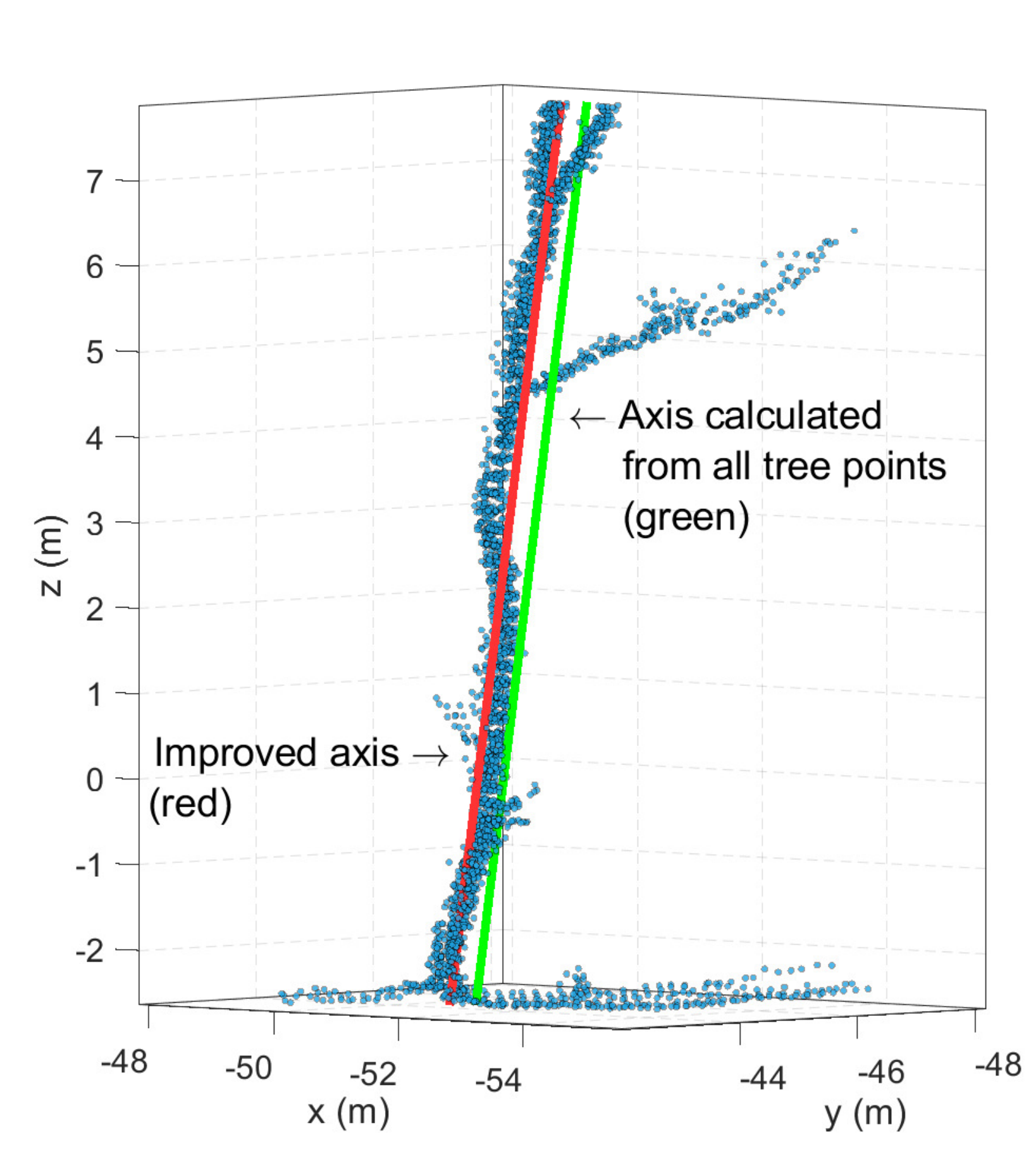}
                \label{fig:improved_lobf}
    \end{subfigure}
    \vspace{-4mm}
    \caption{\small{An example of \textit{Tree Tracking}. \textbf{Left}: The two clusters \(C\) with their
major axis shown and the plane between them. \textbf{Right}: Line of best fit on all tree points
(green) and on the points excluding branch and ground points (red).}}
    \label{fig:tree_tracking}
\end{figure}

Should a cluster satisfy the above properties we deem that it  contains points belonging to a tree. It is then compared to the existing set of confirmed trees \(T\) and merged if a match is found. In
order for a match between two clusters to be found, their major axis must converge to within some
threshold distance at the base of the highest cluster or at a plane segmenting them. This threshold is approximately equal to the maximum tree radius to be considered.
\figref{fig:tree_tracking} (left) demonstrates an example of this.

\algref{alg:tree_tracking} summarises the \textit{Tree Tracking} pipeline.
The evaluation of a match and the subsequent merge process is assessed until no more matches are
found for an input cluster. This is important to ensure all potential clusters belonging to a tree
are merged.

\begin{algorithm}[b]
\caption{ \small{Tree Tracking Algorithm.}}
\label{alg:tree_tracking}
\small{
\SetAlgoLined
\DontPrintSemicolon
\textbf{input}: Set of point clouds $C$ \;
\textbf{output}: Set of updated trees $ T $\;
\Begin{
\For {\textup{Cluster} $ C_i \subset C $}
{
    Calculate the tree descriptor $t_i = f(C_i)$ \;
    \If{$t_i$ is classified as being a tree}
    {
        \eIf {a first match for $t_i$ is found in $ T $ ($T_{j}$) }
        {
            merge cluster into the set of tree points
            $P_{j} = P_{j} \bigcup C_i$ and re-calculate the tree descriptor $t_{j} = f(P_{j}) $ \;
            \While{a subsequent match for the updated $t_{j}$ is found in $ T $ ($T_{tmp}$) }
            {
                merge the two sets of tree points $P_{j} = P_{j} \bigcup P_{tmp}$ and re-calculate the tree descriptor $t_{j} = f(P_{j}) $ \;
                delete $T_{tmp}$ from $T$ \;
            }
        }
        {
        Push back the $t_i,C_i$ into the set $T$
        }
    }
}
}}
\end{algorithm}

\subsubsection{Robust estimation of the tree's major axis}
\label{sec:LoBF}

As mentioned earlier, the calculation of the tree's major axis, or a clusters major axis, is done
using regression on a subset of the tree's points, to find a line of best fit. Beginning with
regression utilising every point in the cluster gives an initial estimate of the incline (\textbf{I}) and its intercept.

However, since the objective is to find the axis of the tree's trunk, least squares regression can
be affected when the branches, ground or crown points are included in the segmentation. Therefore,
points that are a multiple of the standard deviation away from the calculated line or outside the
central 90\% of the tree's height are ignored. The regression is re-calculated with the subset of
points. This is iterated multiple times with successively smaller tolerances on the inclusion of
points. This result is shown in \figref{fig:tree_tracking} (right).


\subsection{Base Segmentation}
\label{sec:Base_estimation}
As the ground has not been removed prior to segmentation, points that are not necessarily portions
of the tree have the potential to be segmented. As a result of this the lowest point in \(P\)
(minimum $z$ value) may not be an accurate estimate of the tree's base. We segment the base as a result.

The two most recent overlapping elevation maps, from \secref{sec:Elevation_Mapping}, are converted into a point cloud so that a search over these `ground points' can be achieved. Points within a 2 metre radius of the lowest point in \(P\) are identified, using a nearest neighbour method from the PCL library. From this set, the closest point to the tree's axis is found and set as the tree's base \textbf{b}.

\subsection{Estimation of Diameter at Breast Height (DBH)}
\label{sec:Estimation_of_DBH}

To estimate the DBH of each individual tree, we employ a  procedure similar to that used in the
literature for general cylinder fitting~\cite{pratt1987direct}. This procedure is broken down into
three parts: firstly, segmenting
the points at breast height, secondly, projection of these points onto a plane, and lastly, fitting
a circle to the projected points.

Utilising the method described by
Zhou~\etal~\cite{zhou2019extracting} and
Heo~\etal~\cite{heo2019estimating}, our system segments the LiDAR points, for recently updated trees in \(T\). Points (\(S\)), from the accumulated point cloud (\(P\)), that are located within a 10 cm height range centered 1.4 m above the tree's base are segmented. These points are projected onto a plane with normal (\(\hat{\textbf{n}}\)) equal to the tree's incline (\(\textbf{I}\)) to ensure the set of points can be modelled closely to a circle. The alternative is to use elliptical fitting for trees angled from the vertical. In practice the sensor only captures data from a single side of the tree, which may cause overfitting when using an elliptical distribution, especially for trees which have irregularly curved surfaces.

The potential problem with this method is that these points are taken from the downsampled point cloud so are likely to be inaccurate. However, an advantage of this method is that points should cover a greater proportion of the tree's surface if the LiDAR sensor measured it from a set of different angles.


RANSAC circle fitting is then used on the set of resultant 2D points, which is robust to
potential outliers from errors in segmentation \cite{Schnabel2007RANSAC}.


\section{Experiments and Evaluation}
\label{sec:Experiments}

\begin{figure}[t]
\includegraphics[width=0.45\textwidth]{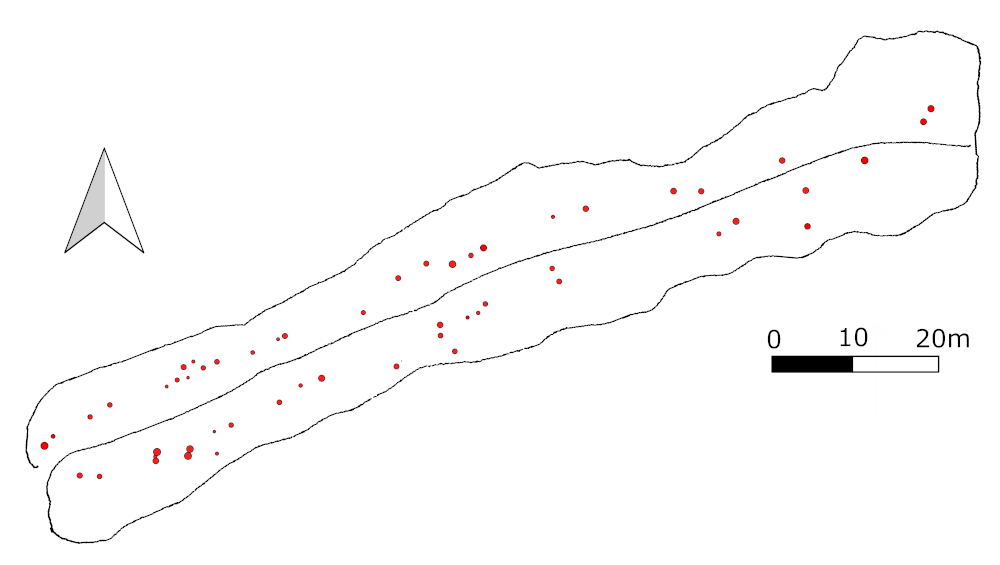}
\caption{\small{Visualisation of the trees selected for measurement within Wytham Woods, Oxford, as
well as
the path taken using the handheld device. The location of the chosen transect is
51$^\circ$46'60.59''N, 001$^\circ$20'21.47''W.}}
\label{fig:map}
\end{figure}



The described system was implemented in C++ and the Robot Operating System (ROS) was used for communication between the hand-held device and the software modules.

Evaluation of the system was carried out using data collected in Wytham Woods, Oxford, UK. The woods is 400 ha in area and has been used for ecology research for over 80 years. It is a Smithsonian ForestGEO site; which is a collection of forestry plots located world-wide used for collective ecology research. Within this scheme, it has taken part in 3 mass censuses and about 16200 trees have been manually measured.

We selected 55 trees along a $\sim$150 m transect. The selected trees had various diameters, ranging
from $\sim$10 cm to $\sim$80 cm, which could be reasonably measured manually at breast height and had clearly visible bases.


We chose a small section of the forest shown in \figref{fig:map}. The DBH of every tree was measured using a tape measure at a height of 1.4 m
from the ground prior to the LiDAR data collection. Two types of LiDAR sensors were
used: The mobile LiDAR device specified earlier and a Leica
BLK terrestrial
LiDAR\footnote{https://leica-geosystems.com/en-gb/products/laser-scanners/scanners/blk360} (for comparison).
The path taken by the mobile LiDAR can be seen in
\figref{fig:map} and \figref{fig:cover_fig}, while a terrestrial LiDAR map was made along the
central path to generate the ground truth for evaluation of our LiDAR odometry system and to provide a performance baseline.

\subsection{LiDAR Odometry}
\begin{figure}[t]
    \centering
        \centering
        \includegraphics[width=0.9\linewidth,trim={1.3cm 0cm 0.5cm 0cm},clip]{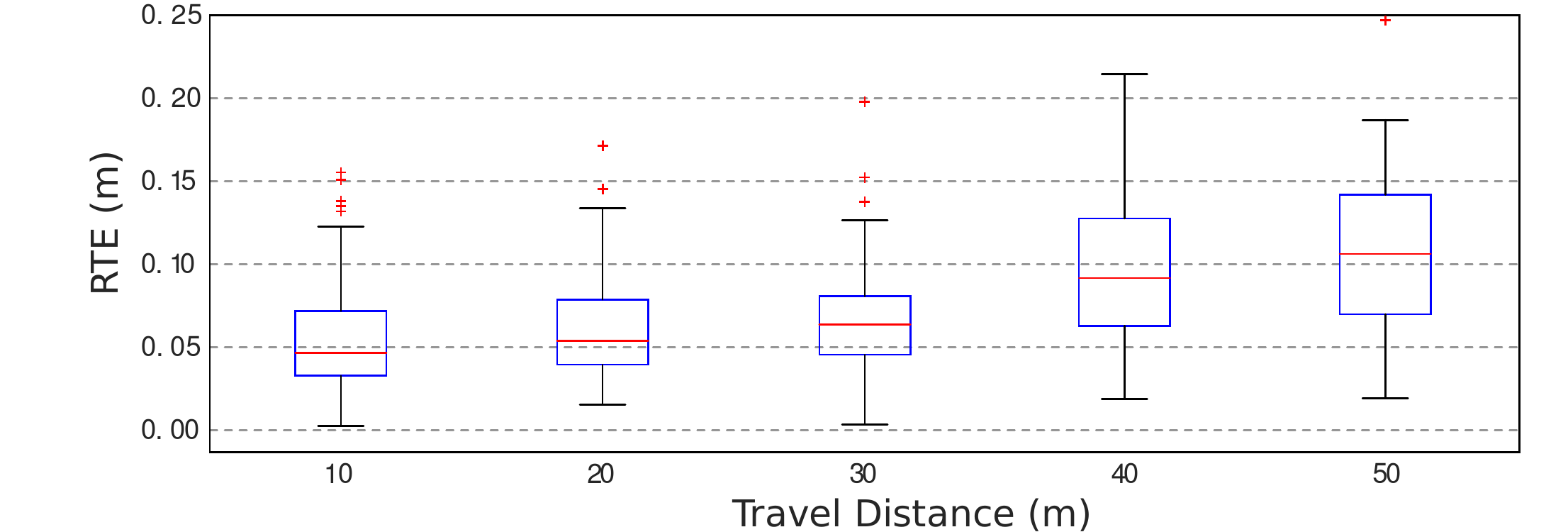}
    \caption{\small{Relative translation error of the LiDAR odometry at different subtrajectory presented as a
series of boxplots. Each box indicates the 25 and 75 percentiles of the estimation errors,
the horizontal line through the box the median, and the whiskers are the maximum and minimum excluding outliers.}}
    \label{fig:odm_eval}
\end{figure}

To evaluate the odometry component of the system, we used the reconstruction created by the Leica
BLK as a prior map.
We then registered the individual point clouds from the hand-held LiDAR against the prior map to generate an accurate
6 degree-of-freedom ground truth trajectory at 10 Hz. The estimated trajectory from the lidar odometry system (without access to the Leica map)
was then
compared against the ground truth trajectory using Relative Translation Error (RTE) at travel
distances of between 10 m and 50 m to produce \figref{fig:odm_eval}.

We aim to accumulate and process LiDAR scans for individual trees as we walk up and pass them at ranges up
to 20 m, thus having a rate of just a few centimetres on this scale is satisfactory.



\label{sec:lidar_odom}


%
%

\subsection{Tree Tracking}

\begin{figure}[b]
    \centering
    \includegraphics[width=1.0\linewidth,trim={5cm 0cm 5cm 1cm},clip]{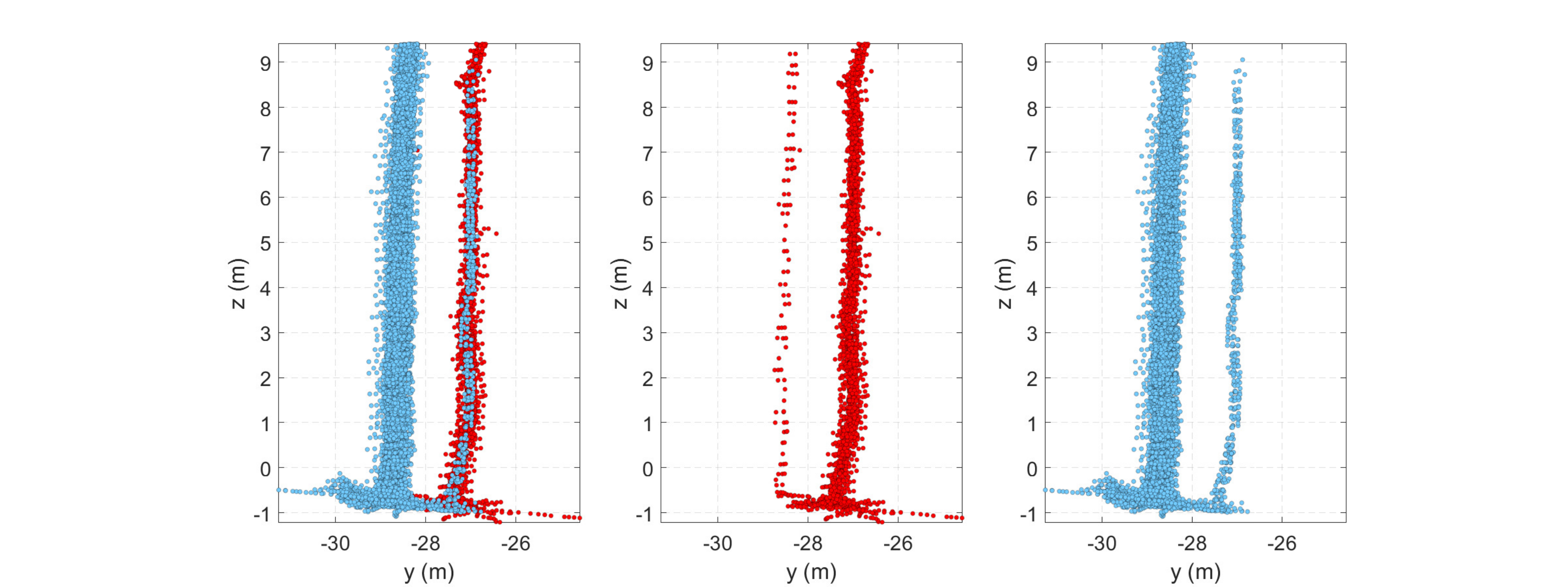}
    \caption{\small{An example of tree tracking failing for trees in close proximity, the ground
points clearly interconnect the two trees. \textbf{Left}: The two trees shown entirely.
\textbf{Center and Right}: Segmented tree with incorrectly assigned points.}}
    \label{fig:unseperated_trees}
\end{figure}

The \textit{Tree Tracking} system performed satisfactorily within the constraints of online performance and
the density of foliage and trees on the test transect. Occasionally, trees were clustered
together when they should not have been, this is most often due to low hanging interconnecting branches
or foliage close to a tree's base. Out of the 55 measured trees 3 of them contained points which belonged to other objects. It was observed that this was also partially due to a transient effect in \textit{Tree Tracking} in which assignment between nearby clusters is uncertain when the tree is first detected. This problem is shown in \figref{fig:unseperated_trees}.

On the other hand, from the 55 measured trees, one failed to be automatically segmented due to its small size, having a DBH $\sim 0.11$ m. This indicates that our current hardware setup and software configuration limits the performance of the system detecting small trees with DBH below 10 cm. 



\subsection{DBH Estimation with LiDAR360}

To provide a baseline, we used the commercial package LiDAR360, in particular its TLS forestry module\footnote{https://greenvalleyintl.com/wp-content/GVITutorials/LiDAR360TLSForest/
LiDAR360TLSForestDataPreprocessing.html}, which is designed for tree segmentation and DBH estimation, on both our handheld LiDAR and Leica BLK datasets.

To evaluate the difference between the estimated values of DBH and the manual ground truth measurements, we use the Root Mean Square Error (RMSE) metric.

LiDAR360 successfully identified 53 of the 55 trees in the Leica BLK data. The remaining two, unidentified trees, were manually selected, then DBH estimation was done, accross all 55 trees, resulting in an overall RMSE of 0.046 m. The mobile LiDAR point clouds were accumulated, after odometry, and also tested using the the same software. Every tree was successfully segmented in the accumulated point cloud and the resulting RMSE was 0.056 m.

\begin{figure}[t]
        \centering
        \includegraphics[width=1.0\linewidth,trim={2cm 0cm 8cm 0.7cm},clip]{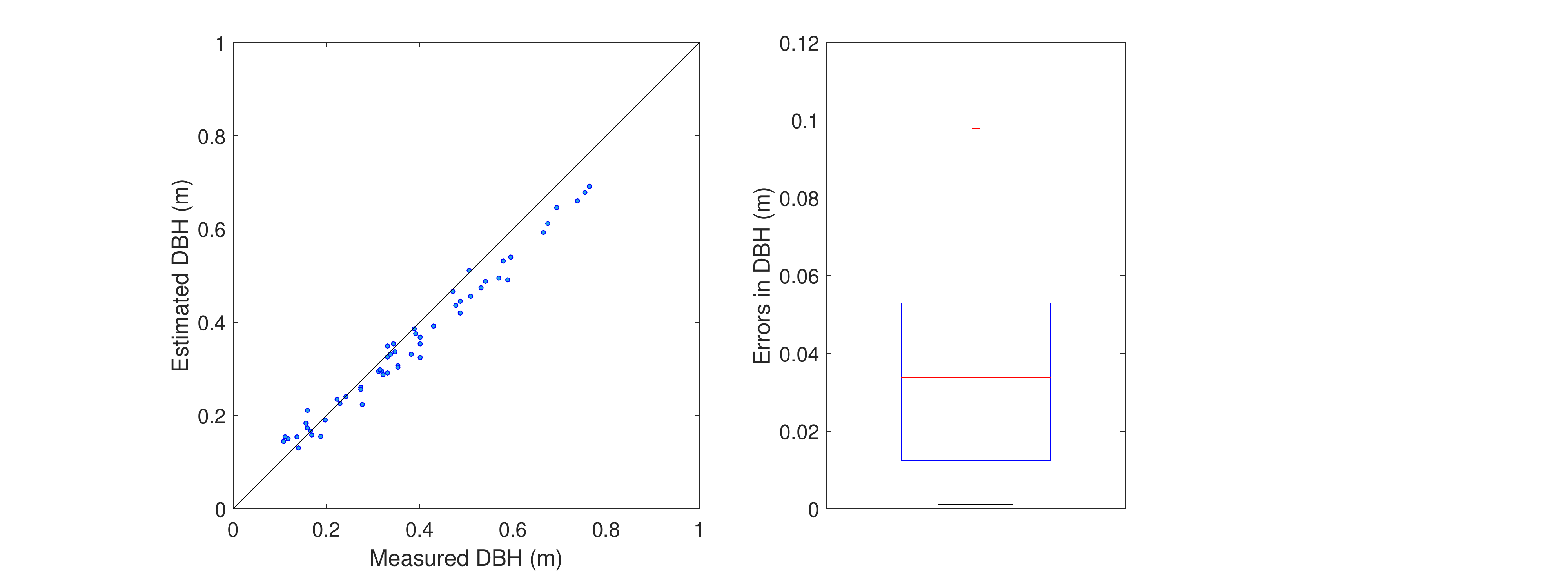}
    \caption{\small{Estimation of DBH using the LiDAR360 Forestry module (using the mobile LiDAR data). \textbf{Left}: Plot of the
55 measured trees vs estimated values as extrapolated from the package. \textbf{Right}: Boxplot of
the errors between the estimated and measured values, RMSE = 0.056 m.}}
    \label{fig:LiDAR360}
\end{figure}

The results for LiDAR360's forestry module are
shown in \figref{fig:LiDAR360}. The graphical results for the terrestrial LiDAR was similar to that of the mobile LiDAR so the graph is omitted.

\subsection{DBH Comparison with Proposed System}

\begin{figure}[t]
    \centering
    \includegraphics[width=1.0\linewidth,trim={2cm 0cm 8cm 0.7cm},clip]{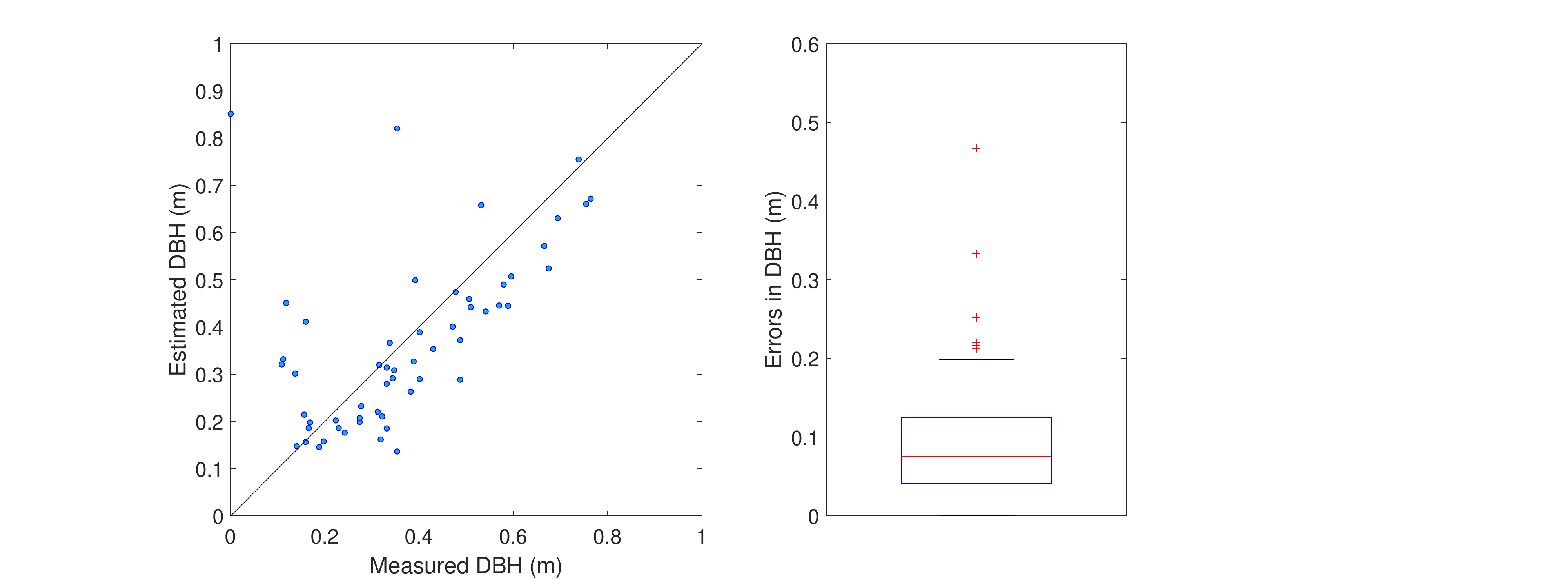}
    \caption{\small{Estimation of DBH using the designed system. \textbf{Left}: Plot of 54 measured
trees vs estimated values, one tree was not segmented. \textbf{Right}: Boxplot of the errors
between the estimated and measured values, RMSE = 0.07 m excluding the outliers and 0.14 m including
them.}}
    \label{fig:ransac}
\end{figure}

\begin{table}[!b]
\centering
\resizebox{\columnwidth}{!}{
\begin{tabular}{cccccccc}
\toprule
\textbf{Algorithm} &Detections &RMSE (m) &Mode &Data used\\
\midrule
\midrule
\textbf{LiDAR360} &53/55 &0.046 &Offline &Leica BLK\\
\textbf{LiDAR360} &55/55 &0.056 &Offline &Ouster (accumulated)\\
\textbf{Proposed} &54/55 &0.07/0.14 &Online &Ouster (live)\\
\bottomrule
\end{tabular}
}
\caption{\small{Summary of the DBH results for the systems evaluated.}}
\label{tab:circle_results}
\end{table}

The results for our system are shown in \figref{fig:ransac}, these results contain data points for every segmented tree (54/55). Three of the trees were poorly segmented which resulted in 
outlier measurements in the figure. The RMSE of these estimates, not including the outliers is 0.07 m.
The RMSE is slightly higher than that of the commercial software, indicating reasonable performance of the system but some room for improvement. Viewing both \figref{fig:LiDAR360} and \figref{fig:ransac} it appears there is a characteristic under estimation using the least squares method, which could be fixed using a fitted model. \tabref{tab:circle_results} summarises the DBH estimates and segmentation results for the proposed system and LiDAR360 using Leica BLK and accumilated mobile LiDAR data.

\subsection{Timing Analysis}
\label{sec:timings}

\begin{figure}[!t]
    \centering
    \includegraphics[width=1.0\linewidth,trim={2cm 0cm 2.5cm 0cm},clip]{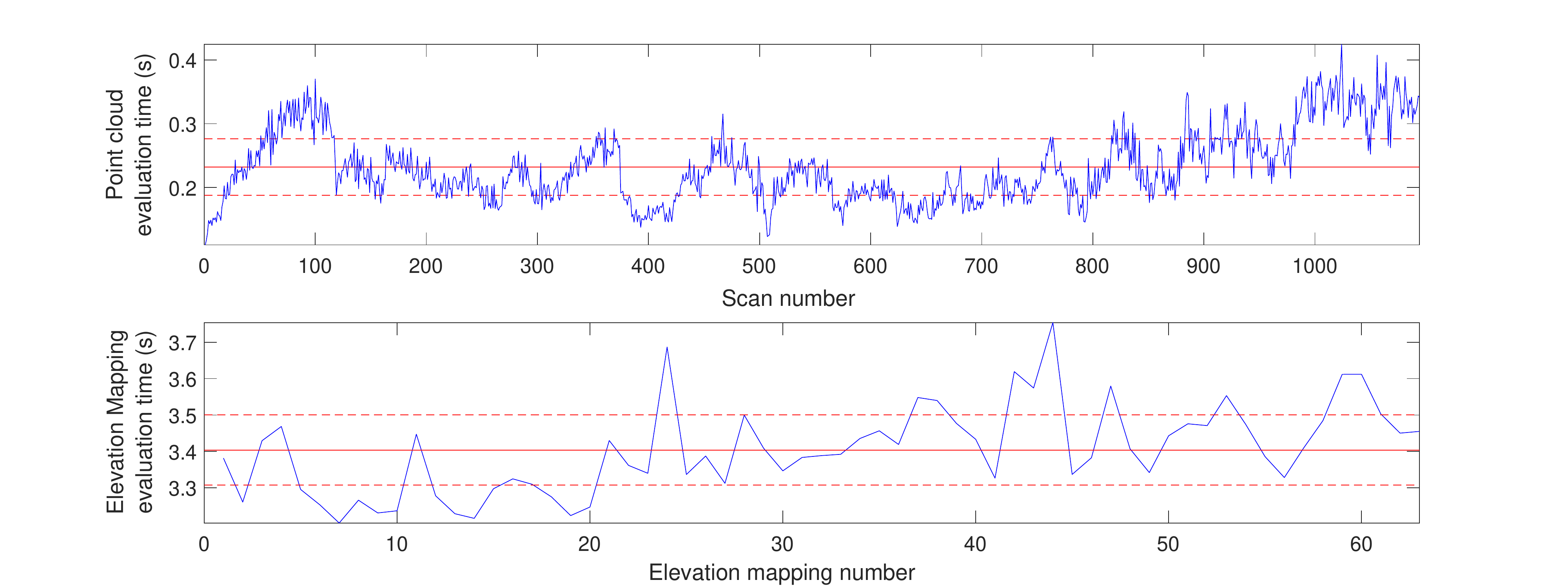}
    \caption{\small{Evolution of computation time, taken over the 622 sec to walk along the Wytham Woods transect, for the two major components. \textbf{Top}: The total time for tree segmentation and fitting. \textbf{Bottom}: Time for the elevation map filtering process (computed every 8 m of travel distance). The mean and standard deviation are shown by the red solid and dashed lines respectively.}}
    \label{fig:Scan_imings}
\end{figure}

To demonstrate that our presented system can run online we carried out an evaluation of the computation time for each component. \figref{fig:Scan_imings} demonstrates the time required for the two major components of the system --- (individual) scan processing in \textit{Tree Tracking} and (accumulated) elevation map filtering. 
A breakdown of individual modules of the \textit{Tree Tracking} pipeline is given in \tabref{tab:timing_specs}. 

The \textit{Tree Tracking} system takes about 0.23 sec (about 4 Hz) to process an individual scan, which is almost two times faster than the output of our 2 Hz LiDAR odometry.

The elevation map filtering process takes 3.4 sec on average and is processed
every $\sim$8 m. Therefore, on average, this
process takes less than half the time available with walking speeds of $\sim$1 m/s. 

Together the two above processes consume less than the available time allowing online operation. 


\section{Conclusion and Future Work}
\label{sec:FutureWork}

This paper presented a mobile LiDAR scanning system for the automatic segmentation of trees and the online estimation of their DBH. The architecture of a tree tracking functional block was introduced to facilitate the
online operation of this system. We tested the system's performance using data from a natural forest, Wytham
Woods, to prove that acceptable results can be achieved online which compares favourably to the performance achieved by a commercial software package conducted in post-processing.


Future work will implement methods within the \textit{Tree Tracking} functional block to overcome long term drift such as loop closure and SLAM pose graph optimisation. 
We would like to consider adaptive segmentation algorithm such as in \cite{digumarti2018automatic} to improve performance on the smaller trees.

Finally we intend to provide feedback to the operator, using the screen, to better direct their actions during operation.

\begin{table}
\centering
\resizebox{\columnwidth}{!}{
\begin{tabular}{cccccccc}
\toprule
\textbf{Processing} &Tree &Circle &Euclidean &Base  &Total
\\
\textbf{Time}&Tracking &Fitting &Clustering &Segmentation &
Time\\
\midrule
\textbf{Mean (ms)} &88 &33 &66 &20 &232\\
\textbf{Std (ms)} & $\pm$30 &$\pm$8& $\pm$33 & $\pm$6  & $\pm$44\\
\bottomrule
\end{tabular}
}
\caption{\small{Computation times for the \textit{Tree Tracking} modules.}}
\label{tab:timing_specs}
\end{table}

\balance
\bibliographystyle{IEEEtran}
\bibliography{library.bib}

\thanks{978-1-6654-1213-1/21/\$31.00 \textcopyright 2021 IEEE}
\end{document}